# Magnetic and the magnetocaloric properties of Ce$_{1-x}$R$_x$Fe$_2$ and Ce(Fe$_{1-x}$M$_x$)$_2$ compounds

Arabinda Haldar[1], K. G. Suresh[1,*] and A. K. Nigam[2]

[1]Department of Physics, Indian Institute of Technology Bombay, Mumbai – 400076, India

[2]Tata Institute of Fundamental Research, Homi Bhabha Road, Mumbai 400005, India

E-mail: suresh@phy.iitb.ac.in

**Abstract.** We have studied selected rare earth doped and transition metal doped CeFe$_2$ compounds by examining their structural, magnetic and magneto-thermal properties. With substitution of Ce by 5 and 10% Gd and 10% Ho, the Curie temperature can be tuned to the range of 267-318 K. Localization of Ce 4$f$ electronic state with rare earth substitutions is attributed for the enhancement of Curie temperature. On the other hand, with Ga and Al substitution at the Fe site, system undergoes paramagnetic to ferromagnetic transition and then to an antiferromagnetic phase on cooling. The magnetocaloric effect across the transitions has been studied from both magnetization isotherms and heat capacity data. It is shown that by choosing the appropriate dopant and its concentration, the magnetocaloric effect around room temperature can be tuned.

## 1. Introduction

Among the $R$Fe$_2$ ($R$ = Rare earth material) series of compounds CeFe$_2$ shows properties which are different compared to the other members of the series [1, 2]. For example, it has anomalous lattice parameter, low Curie temperature ($T_C$ = 230 K) and low saturation magnetization ($M_s$ = $2.4\mu_B$ / $f.u.$). Hybridization between $4f$ - $3d$ orbitals has been shown to be the main reason behind these anomalies [1]. CeFe$_2$ is known to be ferromagnetic (*FM*) with an unstable antiferromagnetic (*AFM*) ground state. Substitution of selected elements [Ru, Re, Ir, Al, Ga, Si etc.] at Fe site stabilizes the fluctuating *AFM* ground state in this material [3]-[8]. Distinct features of first order transition observed across the antiferromagnetic to ferromagnetic transition (on heating) have drawn lots of interest for a long time in these compounds with Fe site substitutions [7]-[11]. Recently we have shown Ga doped compound also can stabilize the antiferromagnetic ground state in CeFe$_2$ compound [7]. Except for a few reports [12], the magnetocaloric properties of Fe site doped CeFe$_2$ compounds are not probed extensively. Such a study is particularly important in view of the multiple magnetic transitions in them. In this paper, we discuss the magnetocaloric properties of Ga/Al doped CeFe$_2$, whose magnetic properties were reported very recently.

Regarding the effect of substitutions at the Ce site, only a very few reports are available in the literature which deal with their magnetic properties and so far there are no reports on their magnetocaloric properties. With rare earth (*R*) substitution, it was shown that one can considerably enhance the $T_C$ but cannot stabilize the *AFM* phase at low temperatures [13]. Therefore, rare earth doped CeFe$_2$ compounds still need to be understood for their fundamental aspects. As $T_C$ can be enhanced with substitution of different rare earths, its magnetocaloric effect is also of interest for investigation. Recent interest in room temperature magnetocaloric effect has given a boost to the research of finding new materials showing large *MCE* with least hysteresis loss. In view of these, we have studied the magnetic and magnetocaloric properties of (Ce$_{1-x}$R$_x$)Fe$_2$ compounds with $R$ = Ho and Gd as well.

The use of a magnetic material as magnetic refrigerant relies on its magnetocaloric behavior. *MCE* is a magneto-thermodynamic phenomenon which gives rise to a change in the temperature caused by a material's exposure to a magnetic field. Magnetic entropy change ($\Delta S_M$) and adiabatic temperature change ($\Delta T_{ad}$) are the measure of *MCE* in a material. It can be measured using magnetization isotherms with the help of the Maxwell's relation [14, 15],

$$\Delta S_M(T, \Delta H) = \int_{H_1}^{H_2} \left( \frac{\delta M(T,H)}{\delta T} \right)_H dH \qquad (1)$$

For *M(H)* isotherms taken at different constant temperatures at discrete temperature intervals, the above relation can be approximated to the following expression [16]:

$$\Delta S_M \approx \frac{1}{\Delta T} \left[ \int_{H_1}^{H_2} M(T+\Delta T, H) dH - \int_{H_1}^{H_2} M(T,H) dH \right] \qquad (2)$$

Magnetocaloric behavior can be well parameterized from heat capacity measurement as a function of temperature in constant magnetic fields, $C(T)_H$. The entropy of a magnetic solid in zero field and in field can be expressed as [14, 15],

$$S(T)_{H=0} = \int_0^T \frac{C(T)_0}{T} dT + S_0$$

and

$$S(T)_{H \neq 0} = \int_0^T \frac{C(T)_H}{T} dT + S_{0,H}, \qquad (3)$$

where $S_0$ and $S_{0,H}$ are the zero temperature entropies in zero field and in presence of a field. In a condensed system these are the same (i.e. $S_0 = S_{0,H}$). Therefore, both $\Delta T_{ad}(T)_{\Delta H}$ and $\Delta S_M(T)_{\Delta H}$ can be calculated as [14, 15],

$$\Delta T_{ad}(T)_{\Delta H} \cong \left[ T(S)_{H \neq 0} - T(S)_{H=0} \right]_S \qquad (4)$$

$$\Delta S_M(T)_{\Delta H} \cong S(T)_{H \neq 0} - S(T)_{H=0} \qquad (5)$$

The quantity that can be used to compare the magnetic refrigeration potential of different materials is the refrigerant capacity or the relative cooling power (*RCP*) which is parameterized as the product of full width at half maximum of $\Delta S_M$ vs. *T* plot and the maximum value of $\Delta S_M$.

## 2. Experimental Details

All the polycrystalline compounds, $CeFe_2$, $Ce_{0.9}Ho_{0.1}Fe_2$, $Ce_{0.95}Gd_{0.05}Fe_2$, $Ce_{0.9}Gd_{0.1}Fe_2$, $Ce(Fe_{0.975}Ga_{0.025})_2$, $Ce(Fe_{0.99}Al_{0.01})_2$ and $Ce(Fe_{0.95}Al_{0.05})_2$ were prepared by arc melting method.

The constituent elements, of at least 99.9% purity, were melted by taking their stoichiometric proportion in a water cooled copper hearth under argon atmosphere. The alloys buttons were remelted several times. The arc melted samples were annealed for 10 days in the following way: 600 ºC for 2 days, 700 ºC for 5 days, 800 ºC for 2 days and 850 ºC for 1 day [3]. The structural analysis was performed by the Rietveld refinement of room temperature x-ray diffraction patterns (*XRD*). Magnetization and heat capacity measurements were performed in the Physical Property Measurement System (PPMS, Quantum Design Model). Magnetization has been measured in zero field cooled (ZFC), field cooled cooling (FCC) and field cooled warming (FCW) modes.

## 3. Results and Discussion

### A. $(Ce_{1-x}R_x)Fe_2$ [R= Gd, Ho] compounds

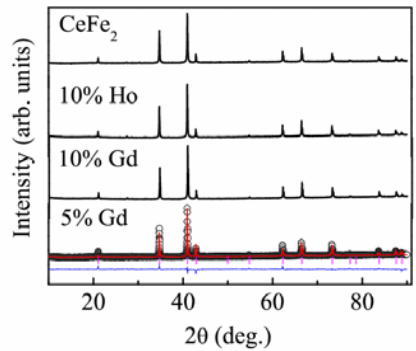

**Figure 1**. Room temperature x-ray diffractograms of $CeFe_2$, $Ce_{0.9}Ho_{0.1}Fe_2$, $Ce_{0.9}Gd_{0.1}Fe_2$ and $Ce_{0.95}Gd_{0.05}Fe_2$ compounds.

Room temperature *X*-ray diffraction patterns are shown in figure 1. Rietveld refinement has been done on all the compounds, but shown only for the 5% Gd doped compound for clarity in view. Refinement shows that all the compounds, like the parent compound, possess the $MgCu_2$ type cubic structure with the space group $Fd\bar{3}m$ [7].

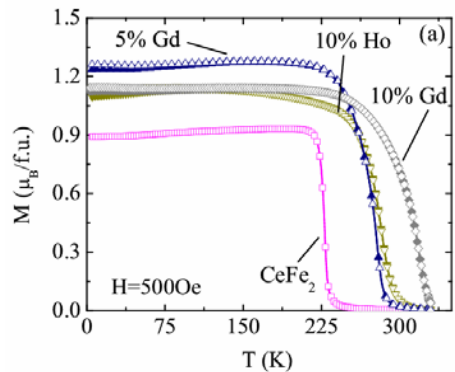

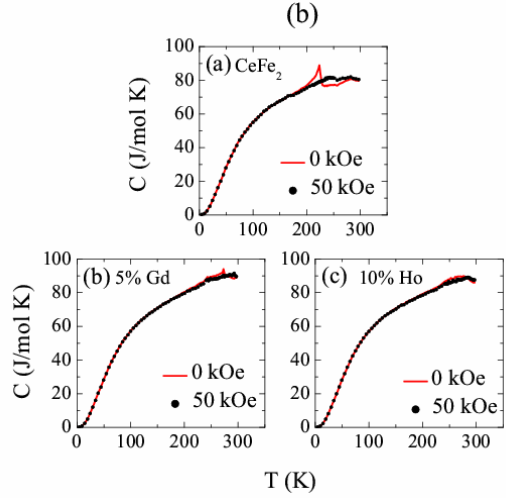

**Figure 2**. (a) Temperature dependence of magnetization of $CeFe_2$, $Ce_{0.9}Ho_{0.1}Fe_2$, $Ce_{0.95}Gd_{0.05}Fe_2$ and $Ce_{0.9}Gd_{0.1}Fe_2$ compounds at H=500 Oe. All the data have been taken during warming the sample both in ZFC (filled symbols) and FCW (open symbols) modes. (b) Heat capacity *vs.* temperature plots for selected compounds in zero and 50 kOe fields. Data has been taken in the ZFC mode.

Temperature variation of magnetization data is shown in figure 2(a) in 2-330 K range with an applied field $H = 500$ Oe. The Curie temperature for the undoped $CeFe_2$ compound is 228 K which is consistent with earlier report [17]. All the compounds show ferromagnetic behavior in the entire temperature range investigated. This is in contrast to the case of substitutions at the Fe site [3, 4, 7]. With substitution of Ce by 5 and 10% Gd and 10% Ho, the Curie temperature increases from 228 K (for $CeFe_2$) to 267-318 K. It is observed that for 5% Gd, $T_C$ is 280 K and for 10% Gd doping, it increases to 318 K. That is, by slightly changing the concentration of rare earths, the $T_C$ can be varied considerably. In $CeFe_2$, Ce 4*f* and Fe 3*d* hybridize and couple ferrimagnetically, which is in contrast to the conventional coupling seen between light rare earth (like Ce) and the transition elements (such as Fe). This is because, though Ce is a rare earth, its 4*f* orbital is more or less itinerant. Substitution of rare earth elements at Ce site causes the localization of the 4*f* electronic state [13]. More localization implies less hybridization between 4*f* and 3*d* states which in turn increases the Fe-Fe direct exchange interaction, which is thought to be responsible for the increase in $T_C$ in the rare earth-transition metal intermetallic compounds.

In figure 2(b) variation of heat capacity data has been shown as a function of temperature, in 3-295 K range. $CeFe_2$ compound shows a sharp peak at $T_C$ which diminishes with 50 kOe magnetic field. $C/T$ *vs.* $T^2$ data at very low temperature show a linear behavior for $CeFe_2$ with an electronic heat capacity coefficient ($\gamma$) value of 47 mJ/mol $K^2$ which is consistent with earlier report [18].

For 5% Gd doped and 10% Ho doped CeFe$_2$ compound, the $\gamma$ values are found to be 44 mJ/mol K$^2$ and 41 mJ/mol K$^2$ respectively. The monotonic decreasing trend in the $\gamma$ value may be due to the increase in the localization of the 4$f$ band. Both the parent and the substituted compounds show a peak near the Curie temperature, indicating the second order nature of the transition. However, it can be seen that the peak gets diminished in the substituted compounds.

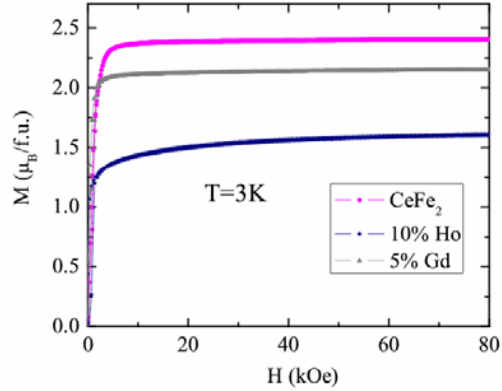

**Figure 3.** Two loop magnetization isotherms at $T$ = 3 K for (Ce, $R$)Fe$_2$ compounds.

Low temperature ($T$ = 3 K), $M$ ($H$) isotherms for the undoped and doped CeFe$_2$ compounds have been plotted in figure 3. The moment values are found to be saturated at 50 kOe. The saturation magnetic moment is found to be 2.4 $\mu_B / f.u.$ for CeFe$_2$ which matches well with earlier reports [1, 2]. All the compounds show ferromagnetic behavior. The saturation moment is found to be 2.4, 2.2 and 1.6 $\mu_B / f.u.$ for CeFe$_2$, 5% Gd and 10% Ho doped CeFe$_2$ compounds respectively. The decrease in the moment with $R$ substitution reflects the fact that the ferrimagnetic coupling between $R$ and Fe sublattices is more in the doped compounds. The order of transition at $T_C$ in these compounds can be probed from the Arrott's plots [7, 19]. The absence of $S$-shaped Arrott's plots across the transition temperature region (not shown) indicates the second order phase transition in all these compounds.

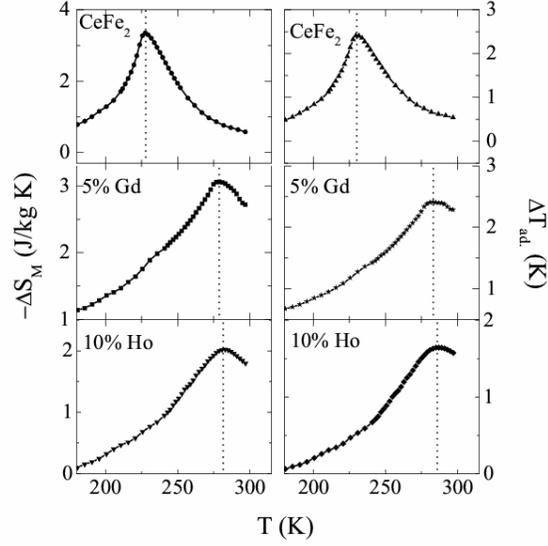

**Figure 4**. (a) Isothermal magnetic entropy change and adiabatic temperature change as a function of temperature in (Ce,*R*)Fe$_2$ compounds for a field change of 50 kOe.

*MCE* has been calculated in terms of isothermal magnetic entropy change ($-\Delta S_M$), using the magnetization isotherms (equation 2) and using heat capacity data (equation 4). In the left panel of figure 4 magnetic entropy change, which was calculated from *C-H-T* data, is shown. It is to be noted here that the $-\Delta S_M$ value is found to be same as obtained from the *M-H-T* data. Substitution of *R* in the place of Ce causes a decrease in $\Delta S_M$ values. However, the peak becomes broader, which is necessary for a good refrigerant material. Moreover, different *R* substitutions give a *MCE* peak over a broad temperature region near the room temperature, without much change in the peak value. The values of maximum entropy change ($-\Delta S_M^{max}$) for $\Delta H = 50 kOe$ have been listed in table 1. Adiabatic temperature change, $\Delta T_{ad}$, has been calculated (equation 5) from the *C-H-T* data and is shown in right panel of figure 4. For CeFe$_2$ the maximum adiabatic temperature change ($\Delta T_{ad}^{max}$) is observed to be 2.4 K at 230 K. It is noteworthy here that with 10% Ho substitution, the $T_C$ has increased to 283 K, but the $\Delta T_{ad}^{max}$ remains unchanged at 2.4 K. For Gd doped compound this value is slightly less than 2 K. Another parameter of interest for magnetic refrigerant materials is the low hysteresis near the *MCE* peak region. *M(H)* isotherms has been taken during increasing and decreasing fields to calculate the hysteresis loss, which turns out to be almost zero in all the compounds. Thee quality factor of a refrigerant material is the refrigeration capacity or relative cooling power (*RCP*). *RCP* for these compounds has been calculated by taking the product of maximum entropy change and full width at half maximum (*FWHM*) of $\Delta S_M$ *vs. T* curves [20] as discussed in the introduction section. $\Delta S_M$ *vs. T* plot was extrapolated to draw a full envelope around the peak value of $\Delta S_M$ for calculating *FWHM*. *RCP* values thus calculated are tabulated in Table 1. The increase in the FWHM is due to the random distribution of the substituted atoms, which produces a distribution of magnetic transition temperatures.

**Table 1.** Curie temperatures ($T_C$), maximum change of entropy $(-\Delta S_M^{max})$, maximum adiabatic temperature change $(\Delta T_{ad}^{max})$ and relative cooling power (RCP) in CeFe$_2$ and rare-earth doped CeFe$_2$ compounds, for a field change of 50 kOe.

| Compound | $T_C$ (K) | $-\Delta S_M^{max}(Jkg^{-1}K^{-1})$ ($\Delta H = 50 kOe$) | $\Delta T_{ad}^{max}(K)$ ($\Delta H = 50 kOe$) | RCP (J kg$^{-1}$) |
|---|---|---|---|---|
| CeFe$_2$ | 228 | 3.8 | 2.4 | 141 |
| Ce$_{0.9}$Ho$_{0.10}$Fe$_2$ | 283 | 1.9 | 2.4 | 112 |
| Ce$_{0.95}$Gd$_{0.05}$Fe$_2$ | 280 | 2.1 | 1.6 | 120 |
| Ce$_{0.9}$Gd$_{0.10}$Fe$_2$ | 318 | 1.6 | * | 108 |

* Not measured as the $T_C$ is close to the safe upper limit of the variable temperature insert of PPMS.

**B. Ce(Fe$_{1-x}$M$_x$)$_2$ [M = Ga, Al] compounds:**

Although '*R*' substitution at Ce site cannot stabilize the low temperature *AFM* state, certain substitution at Fe site stabilizes it. This gives rise to *FM*-paramagnetic (*PM*) and *AFM-FM* transitions in such materials. We use Ga/Al substituted compounds namely, Ce(Fe$_{0.975}$Ga$_{0.025}$)$_2$, Ce(Fe$_{0.99}$Al$_{0.01}$)$_2$ and Ce(Fe$_{0.95}$Al$_{0.05}$)$_2$ to study the *MCE* variation across these transitions. We have recently shown that Ga substituted CeFe$_2$ compound undergoes an antiferromagnetic transition below $T_C$ [7]. It was also found that the $T_C$ decreases with increase in Ga concentration.

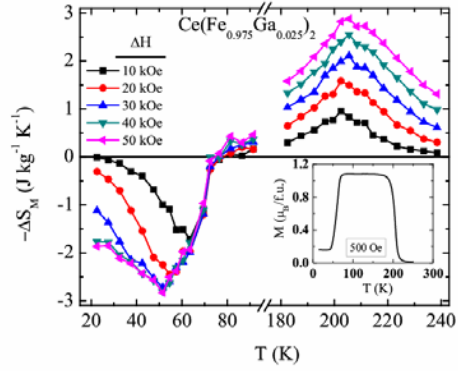

**Figure 5.** Isothermal negative magnetic entropy change ($-\Delta S_M$) as a function of temperature in Ce(Fe$_{0.975}$Ga$_{0.025}$)$_2$ compound. The inset shows the *M-T* data in 500 Oe field.

Temperature dependence of magnetization data shown in the inset of figure 5 shows that Ce(Fe$_{0.975}$Ga$_{0.025}$)$_2$ undergoes *PM* to *FM* transition at $T_C$ = 206 K and *FM* to *AFM* phase at the Neel temperature ($T_N$) = 64 K. We have calculated the magnetic entropy change using magnetization isotherms across both these transition regions in this compound, as shown in figure 5. Interestingly the sign of the *MCE* is different in the two transition regions, resulting in an oscillatory *MCE* behavior. This is expected because of the fact that the transitions are *AFM-FM* and *FM-PM* in nature. The Entropy change is positive across the *AFM* to *FM* transition region and it is negative across *FM* to *PM* transition region. The $-\Delta S_M$ peak across *AFM-FM* transition region is found to shift towards lower temperature considerably with increase in field. This is a reflection of the decrease in the Neel temperature with the increase in the field, as usually seen in antiferromagnets. As maximum entropy change occurs at the transition region *MCE* peak shifts to lower temperatures. The entropy change values are tabulated in Table 2. The *RCP* in this case is found to be more than that of Ru doped CeFe$_2$ compound [12]. Across *PM-FM* (*FM-AFM*) transition *RCP* value is found to be 145 J/kg (127 J/kg) which is 79.4 J/kg (59.9 J/kg) in Ce(Fe$_{0.96}$Ru$_{0.04}$)$_2$ compound [12]. This difference is expected as the transition is broad in the case of Ga doping compared to Ru doping. It may also be noted that the *MCE* values are nearly the same in both *R* doped CeFe$_2$ as well as in Ga doped CeFe$_2$.

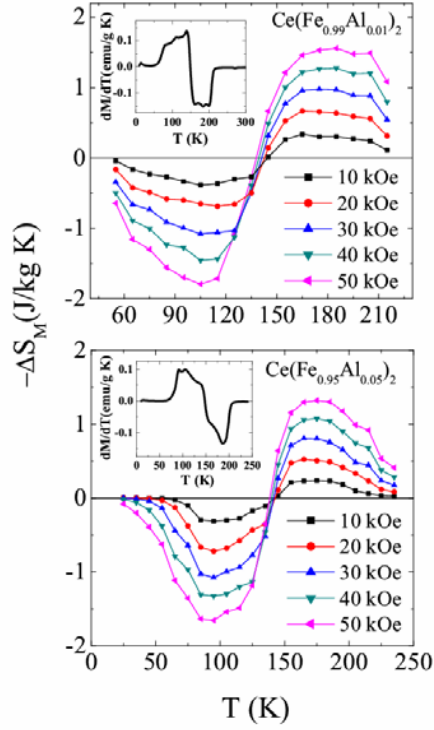

**Figure 6.** ($-\Delta S_M$) *vs.* temperature for $Ce(Fe_{0.99}Al_{0.01})_2$ and $Ce(Fe_{0.95}Al_{0.05})_2$ compounds. Insets show *dM/dT vs. T* plot for both the compounds.

As mentioned earlier, Al is another substituent which causes the low temperature *FM-AFM* transition in $CeFe_2$. Therefore, we have calculated the magnetic entropy change in two selected compounds namely $Ce(Fe_{0.99}Al_{0.01})_2$ and $Ce(Fe_{0.95}Al_{0.05})_2$ on which magnetic properties under pressure have been discussed earlier [21]. The *AFM-FM* and the *FM-PM* transitions for these two compounds are quite broad in this Al doped compounds which are shown in the *dM/dT vs. T* plots in the insets of figure 6. Here also the oscillatory *MCE* behavior is seen from figure 6. Increase in Al concentration has an effect in reducing the magnetic moment as well as the entropy value. The smaller values of $-\Delta S_M^{max}$ compared to Ga doped compound may come due to the experimental protocol difference, as in this case measurement has been taken comparatively large temperature step resulting the reduction of $-\Delta S_M^{max}$. In the Al doped compound also the transitions are broad, which results in comparatively large *RCP* value although $-\Delta S_M^{max}$ is smaller. Table 2 summarizes the *MCE* values in Ga and Al doped compounds, along with the transition temperatures.

As can be seen from the above results, magnetocaloric effect is strongly dependent on the type of magnetic transition. The main difference between Ga and Al doped $CeFe_2$ compounds is that in the former case, the transition is sharper compared to that of the latter. A clear reflection of this is observed in the magnetocaloric effect, resulting in a rather broad peak in Al doped compounds compared to the Ga doped compound. Comparing our results with Ce(Fe$_{0.96}$Ru$_{0.04}$)$_2$, we find that though the *MCE* value is comparatively small, the *RCP* value is large in the present case [12]. This may be attributed to the broad magnetic phase transitions in these compounds.

**Table 2.** The maximum change of entropy $(-\Delta S_M^{\max})$ and the relative cooling power (*RCP*) in Ga and Al doped CeFe$_2$ compounds across *FM-AFM* and *PM-FM* transition regions, for a field change of 50 kOe. The quantities in brackets show the uncertainty in the data.

| Compound | $T_N$(K) | $T_C$ (K) | $-\Delta S_M^{\max}(Jkg^{-1}K^{-1})$ | | *RCP* (J/kg) | |
|---|---|---|---|---|---|---|
| | | | *PM-FM* | *FM-AFM* | *PM-FM* | *FM-AFM* |
| CeFe$_2$ | * | 230 | 3.8 | * | 141 | * |
| Ce(Fe$_{0.975}$Ga$_{0.025}$)$_2$ | 64 | 206 | 2.8 | -2.9 | 145 | 127 |
| Ce(Fe$_{0.99}$Al$_{0.01}$)$_2$ | 108(5) | 178(5) | 1.5 | -1.8 | 103 | 113 |
| Ce(Fe$_{0.95}$Al$_{0.05}$)$_2$ | 115(5) | 175(5) | 1.3 | -1.6 | 98 | 112 |

*No *FM-AFM* transition is observed in this case.

## 4. Conclusions

We have shown that by substitution of Ce with some selected rare earth elements, one can tune the $T_C$ towards room temperature. The *MCE* across the transition region has been studied using the magnetization isotherms and heat capacity data. The $\Delta S_M^{\max}$ value is found to decrease with rare earth substitution, but controlled tuning of the *MCE* peak can be achieved by suitably fixing the rare earth and its concentration. Substitution at the Fe site by Ga and Al causes two transitions namely *PM-FM* and *FM-AFM*. Across these two transition regions, $\Delta S_M$ shows sign reversal. Broad maxima around the transition temperatures result in larger *RCP* values. In both the Ce and the Fe site substituted compounds, the magnetocaloric properties seem to be strongly correlated with the magnetic properties. Though the *MCE* values achieved in this work are not sufficient for commercial applications, the tunability of the *MCE* and the underlying physics are of importance in the design of novel and potential magnetic refrigerant materials for room temperature applications.

*Acknowledgements*

KGS and AKN thank BRNS for the financial support for carrying out this work. The authors also thank D. Buddhikot for his help in certain measurements.